\documentclass[sigconf]{acmart}
\usepackage{amsmath}
\usepackage{algorithm}     
\usepackage{algpseudocode}
\usepackage{enumitem}
\usepackage{float}
\usepackage{booktabs} 
\usepackage{multirow}

\usepackage{booktabs} 
\usepackage{colortbl} 
\usepackage{xcolor}   
\usepackage{threeparttable}
\usepackage{caption}

\usepackage{adjustbox}
\usepackage{graphicx}
\usepackage[table,xcdraw]{xcolor}
\usepackage{amsmath}
\usepackage{caption}
\usepackage{makecell}
\usepackage{enumitem}
\usepackage{booktabs}

\usepackage{tabularx}
\usepackage{array}

\AtBeginDocument{%
  }

\setcopyright{acmlicensed}
\copyrightyear{2018}
\acmYear{2018}
\acmDOI{XXXXXXX.XXXXXXX}
\acmConference[Conference acronym 'XX]{Make sure to enter the correct
  conference title from your rights confirmation email}{June 03--05,
  2018}{Woodstock, NY}
\acmISBN{978-1-4503-XXXX-X/2018/06}

\begin{document}

\title{MMP-Refer: Multimodal Path Retrieval-augmented LLMs For Explainable Recommendation}

\author{Xiangchen Pan}
\affiliation{%
  \institution{Huazhong University of Science and Technology}
  \city{Wuhan}
  \country{China}}
\email{pxcstart666@gmail.com}

\author{WeiWei}
\authornote{Corresponding author.}
\affiliation{%
  \institution{Huazhong University of Science and Technology}
  \city{Wuhan}
  \country{China}
}
\email{weiw@hust.edu.cn}

\renewcommand{\shortauthors}{Trovato et al.}

\begin{abstract}
Explainable recommendations help improve the transparency and credibility of recommendation systems, and play an important role in personalized recommendation scenarios. At present, methods for explainable recommendation based on large language models(LLMs) often consider introducing collaborative information to enhance the personalization and accuracy of the model, but ignore the multimodal information in the recommendation dataset; In addition, collaborative information needs to be aligned with the semantic space of LLM. Introducing collaborative signals through retrieval paths is a good choice, but most of the existing retrieval path collection schemes use the existing Explainable GNN algorithms. Although these methods are effective, they are relatively unexplainable and not be suitable for the recommendation field.


To address the above challenges, we propose MMP-Refer, a framework using \textbf{M}ulti\textbf{M}odal Retrieval \textbf{P}aths with \textbf{Re}trieval-augmented LLM \textbf{F}or \textbf{E}xplainable \textbf{R}ecommendation. We use a sequential recommendation model based on joint residual coding to obtain multimodal embeddings, and design a heuristic search algorithm to obtain retrieval paths by multimodal embeddings; In the generation phase, we integrated a trainable lightweight collaborative adapter to map the graph encoding of interaction subgraphs to the semantic space of the LLM, as soft prompts to enhance the understanding of interaction information by the LLM. Extensive experiments have demonstrated the effectiveness of our approach. Codes and data are available at https://github.com/pxcstart/MMP-Refer.
\end{abstract}

\keywords{Explainable Recommendation, MultiModal Recommendation, Large Language Model, Retrieval-Augmented Generation}


\maketitle



\section{Introduction}
Recommendation system is very important to effectively alleviate the problem of user information overload~\cite{luo2022hysage,luo2022personalized, luo2022towards, luo2024perfedrec++}. In recent years, explainable recommendation has attracted more and more attention~\cite{peake2018explanation, zhang2020explainable, zhang2014explicit}. Through the reasoning of potential causes behind the historical interaction between users and items, the recommendation model can provide explainable suggestions and insights into the decision-making process, so as to make more personalized and transparent recommendations. This not only enhances the model's understanding of user preferences, but also enhances the credibility of the model.

Early research on explainable recommendations mainly used deep learning techniques to generate explanations by utilizing review information, user and item IDs, and user features. For example, Att2Seq~\cite{dong2017learning} and NRT~\cite{li2017neural} use attention mechanisms and recurrent neural networks (RNNs) to generate text explanations, while PETER further explores the use of Transformers~\cite{li2021personalized, li2023personalized} in text generation... Although these methods can generate richer and smoother text explanations compared to early methods based on predefined sentence templates~\cite{zhang2014explicit,li2021caesar}, these methods heavily rely on ID embedding and have limited generalization ability. In recent years, with the rise of large language models(LLMs), they have gained widespread attention in explainable recommendation tasks due to their excellent language generation and comprehension abilities~\cite{chen2024graphwiz,li2024graph,li2023survey}.

In order to provide more personalized explanations, some scholars have attempted to introduce collaborative information into LLMs for generating more accurate and informative explanations. For example, XRec~\cite{ma2024xrec} designed a collaborative instruction to inherit collaborative signals and perform lightweight fine-tuning on LLM; Considering the gap between collaborative information and LLM semantic space, G-Refer~\cite{li2025g} captured collaborative information through node-level and path-level methods, and retrieved corresponding knowledge by searching retrieval paths on the interaction graph. Although the effect is significant, there is still space for improvement: (1) These methods only consider how to introduce collaborative signals, but ignore the multimodal information in the dataset. However, balancing modality gaps and maintaining the uniqueness of specific modalities is a challenge. (2) At present, most of the existing retrieval path collection schemes use Explainable GNN algorithms. Although these methods are effective, they lack explainability and may not be suitable for the recommendation field. Driven by these challenges, we propose a LLM-based explainable recommendation framework based on multimodal retrieval paths.

In MMP-Refer, we adopt a two-stage paradigm of retrieval generation for explainable recommendation. In the retrieval stage, We first obtain multimodal representations of the user side and item side through user behavior sequential modeling based on joint residual encoding, and collect retrieval paths through heuristic search based on the multimodal representations; In the generation stage, we use instruction tuning for the LLM from two levels: path level and graph level, so that the LLM can better understand user preferences. For path level, we will retrieve the profile information of each node in the path in the form of a text sequence as a hard prompt for instructions; For graph level, we encode the retrieval subgraphs corresponding to the retrieval path into graphs, and then map them to the LLM semantic space as instruction soft prompts through a learnable lightweight collaborative adapter.

\begin{figure}[t]
    \centering
    \includegraphics[width=1\linewidth]{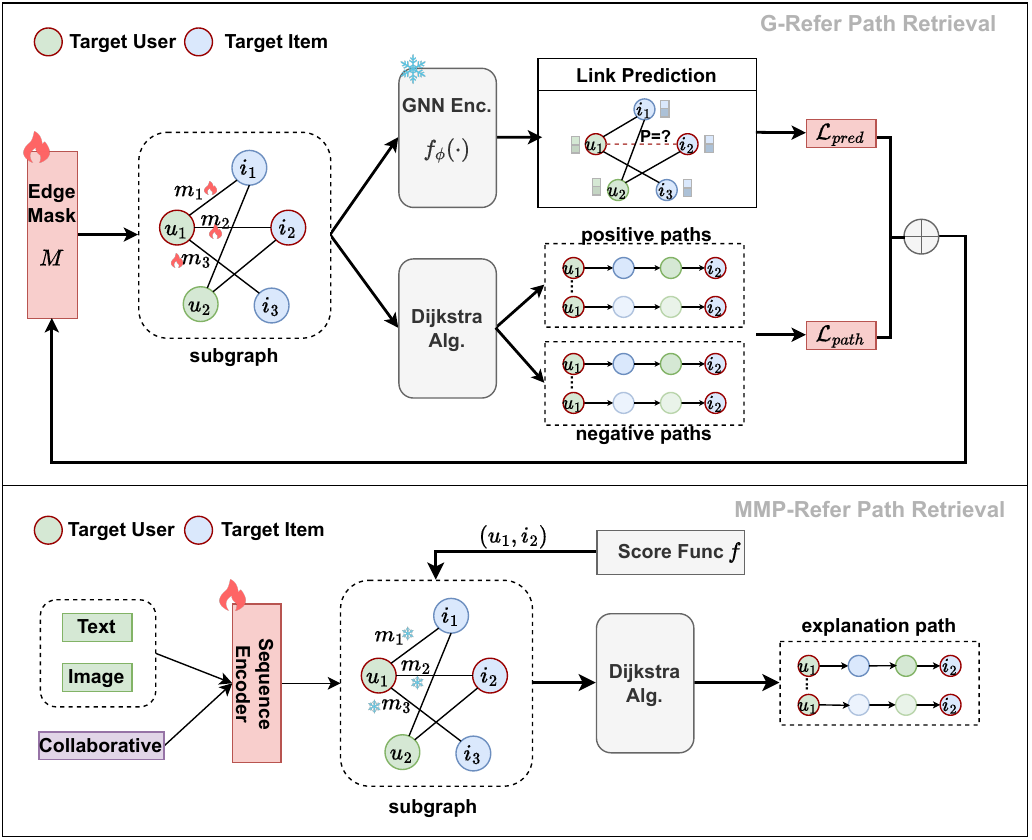}
    \caption{Comparison of retrieval path collection methods between G-Refer and Ours(MMP-Refer).}
    \label{intro}
\end{figure}
 
To address the issue of how to perform multimodal representation and alleviate the information inconsistency caused by Modality Gap, we adopted a training method based on a joint residual quantization encoder to quantitatively encode the item features of different modalities. For user side multimodal representation learning, we use traditional sequential recommendation models to fuse multimodal information and behavioral information together to refine user preference representation. Compared to the fusion method of simply concatenating various modalities, our method can better bridge the modality gap and fit user preferences.

To address the issue of how to efficiently collect explainable retrieval paths, we propose a heuristic search algorithm based on multimodal representation. Considering the effects of recommendation noise and popularity bias, we assume that a retrieval path with the shortest possible hop count, the lowest possible node in degree, and multimodal features as close as possible to the analyzed node is a high-quality explainable path. We will use node representations based on rules to construct weights for edges in each interaction subgraph. As shown in Fig. ~\ref{intro}, compared to the GNN Explanation method that uses edge weights as learnable parameters, our method has higher stability and stronger explainability in the recommendation domain.

The contributions of this work are summarized below:
\begin{itemize}[leftmargin=*] 
    \item We propose a framework MMP-Refer designed to use Multimodal Path Retrieval-augmented LLMs For Explainable Recommendation. By modeling user sequences based on joint residual encoding, multimodal representations of the user side and item side are obtained, we can obtain more more personalized and accurate multimodal representation for explanation.
    \item We propose a heuristic search algorithm for efficiently retrieving explainable paths for specific user-item pairs. Compared to Explainable GNN methods, our method demonstrates stronger adaptability to recommendation data and achieves better retrieval performance.
    \item We conducted extensive experiments on three datasets, and the experiment results validate the effectiveness and flexibility of the model.
\end{itemize}

\section{Related Work}
\subsection{Explanation Recommendation}
Explainable recommendation~\cite{chang2023knowledge,chang2024path,peake2018explanation} has the reasoning ability to explore user recommendation behavior logic, thereby enhancing the personalization and transparency of recommendation systems, which has significant research significance. Early methods mainly used pre-defined templates to generate explanations~\cite{li2021caesar} or extract logical reasoning rules from recommendation models~\cite{chen2021neural,zhu2021faithfully}. With the development of deep learning technology, many works adopt attention mechanisms and recursive neural networks to generate natural language explanations~\cite{dong2017learning, li2017neural}. Currently, the most popular approach is to use LLMs for explainable recommendations. Researchers attempt to generate smoother and more readable explanations by utilizing textual information of users and items. Recently, some studies have attempted to incorporate behavioral information from graphs into LLM to enhance the explanation generated by LLM. For example, XRec~\cite{ma2024xrec} uses Graph Neural Networks(GNNs) to generate behavioral features and maps collaborative embeddings to the LLM semantic space to generate interpretations; G-Refer~\cite{li2025g} introduces the concept of retrieval path based on X-Rec, making collaborative information integration more fine-grained and further enhancing the credibility of explainable suggestions. However, these methods all ignore the multimodal information of the recommendation dataset. In addition, how to make the retrieval path as concise and explainable as possible is also a key challenge. 

\begin{figure*}[t]
    \centering
    \includegraphics[width=1\linewidth]{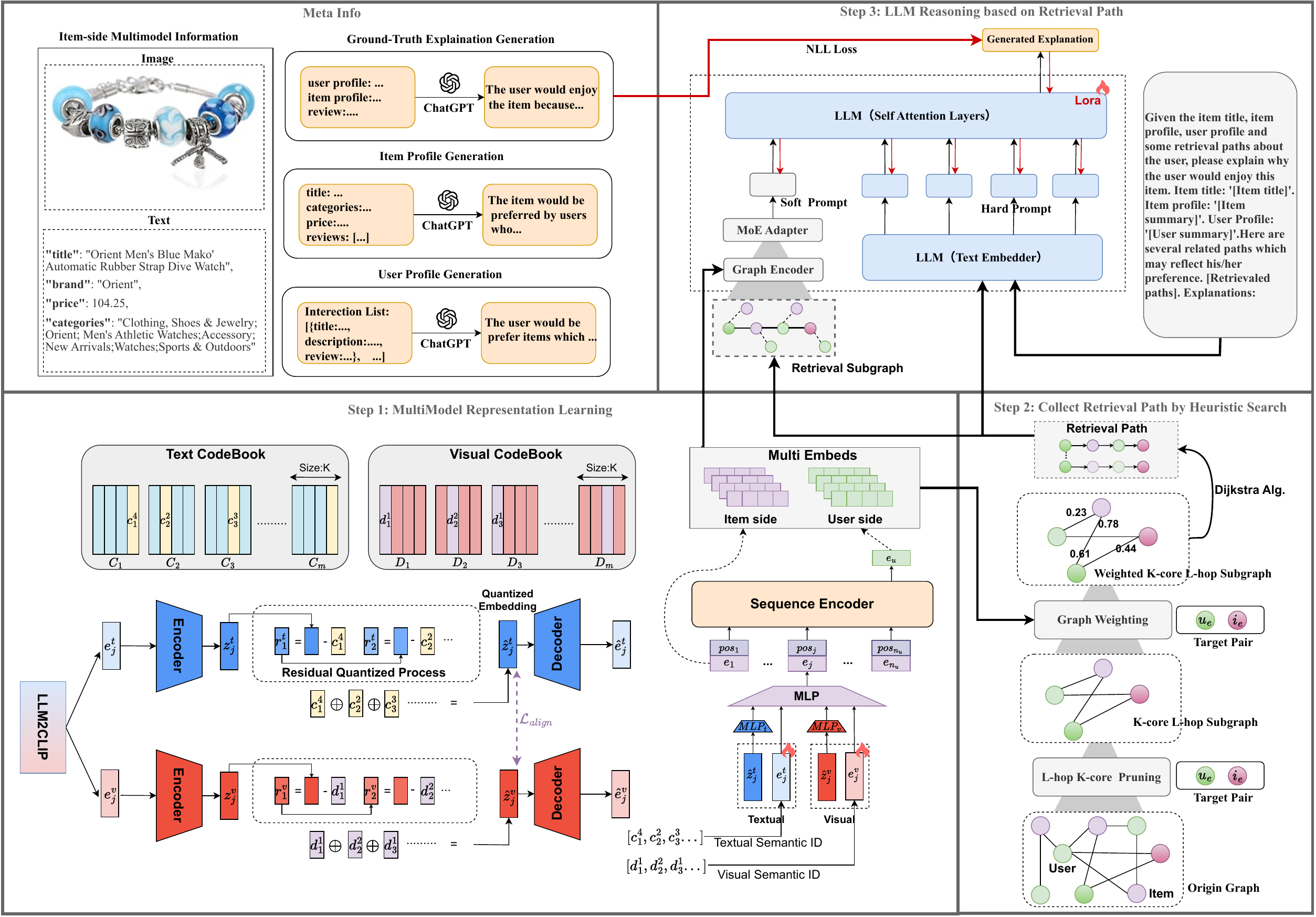}
    \caption{The overview framework of MMP-Refer.It mainly consists of three modules. Firstly, the multimodal representation learning module obtains the multimodal representations of users and items. Then, based on the multimodal features, rules are formulated and retrieval paths are collected through heuristic search; Finally, the textual features and collaborative features of the retrieval path are extracted as external knowledge to assist in fine-tuning LLM for generating high-quality recommendation explanations.
}
    \label{overview}
\end{figure*}

\subsection{GNN Explanation}
GNN Explanation mainly used graph neural networks~\cite{kipf2016semi,velivckovic2017graph,xu2018powerful} to explain the logic behind node and graph classification, where explanation is defined as an important subgraph. There are significant differences in the definition of importance and subgraph selection methods among existing methods. GNNExplainer~\cite{ying2019gnnexplainer} selects edge-induced subgraphs by learning fully parameterized masks on graph edges and node features, where the mutual information (MI) between the masked graph and the predictions made using the original graph is maximized. PGExplainer~\cite{luo2020parameterized} uses the same MI importance, but trains a mask predictor to generate discrete masks. Another popular measure of importance is game theory values. SubgraphX~\cite{yuan2021explainability} uses Shapley values~\cite{shapley1953value} and performs Monte Carlo Tree Search(MCTS) on subgraphs. GStarX~\cite{zhang2022gstarx} uses structure-aware HN values~\cite{hamiache2020associated} to measure the importance of nodes and generate subgraphs induced by important nodes. From other perspectives, there are more studies that are less relevant to this work, such as surrogate models~\cite{huang2022graphlime,vu2020pgm}, counterfactual explanations~\cite{lucic2022cf}, and causal relationships~\cite{lin2021generative,lin2022orphicx}. Although these methods generate subgraphs as explanations, what constitutes a good explanation is a complex topic. In this work, we need to find a highly explainable retrieval path from the interaction graph as external knowledge for LLM inference. The above methods cannot adapt well to the recommendation scenario. We designed a heuristic search method that sets edge weights for interaction graphs through rules, rather than training edge weights as learnable weights.

\subsection{Mutlimodal Recommendation}
Multimodal information is widely present in the interaction between recommendation systems and users, playing a crucial role in user decision-making. In recent years, various works have explored incorporating multimodal information into user preference modeling. Early methods introduced modality information as auxiliary features, extracted them through pre-trained neural networks, and then fused multimodal features with behavioral features using contrastive learning to better simulate user preferences ~\cite{he2016vbpr,wang2017your,xu2021multi,cui2018mv}. These methods usually ignore the problem of information imbalance caused by modality gaps. Simple modality fusion may lead to the model relying more on a certain mode, affecting the generalization ability. In this work, we proposes a user sequential modeling scheme based on joint residual encoding, which is used to better bridge the modality gap and fit user preferences while performing multimodal representation.
\section{Methodology}
In this chapter, we will provide a detailed introduction on how to implement a LLM-based explainable recommendation based on multimodal retrieval paths. The model framework is shown in Fig. ~\ref{overview}. As can be seen, the model mainly consists of three important modules: Multimodal representation learning module; Retrieval path collection module; Large language model inference module. We will provide clear task definitions and implementation of each module in the following subsections.


\subsection{Overview}
\textbf{Task definition:} Explainable recommendation aims to infer the reasons behind recommendation behavior and generate a logically clear textual explanation. In the recommendation scenario, assuming that the user set and item set are defined as $U=\{u_1, u_2,..., u_{|U|}\}$, $V=\{v1,v2,...,v_{|V|}\}$, where each user $u_i$ and item $v_j$ has corresponding text summary information $b_i \in \mathcal{B}$ and $c_j \in \mathcal{C}$ used to summarize user interests and item attributes, respectively. The interaction between users and items can be modeled as a bipartite graph $G=\{(u, v) | u \in U, v \in V\}$, where each edge $(u_i, v_j)$ of graph $G$ represents a recommendation behavior, i.e., user $u_i$ likes item $v_j$. Therefore, we formalize the explainable recommendation task as, for a given recommendation behavior $(u_i, v_j)$, using the interaction graph $G$ and configuration information $b_i$, $c_j$ to analyze and summarize why user $u_i$ likes item $v_j$ by LLM:
\begin{equation}
\begin{aligned}
explanation(u_i,v_j)=LLM(u_i,v_j,G,b_i,c_j)
\end{aligned}
\end{equation}

\textbf{Retrieval path:} In order to explore user preferences more deeply from recommended behaviors and provide explanations, we collect retrieval paths related to recommended behaviors $(u_i, v_j)$ based on the interaction graph $G$, and use them to assist the LLM in inferring and generating explanations. In order to make the retrieval path shorter and more interpretable, we first prune the original graph to obtain the interaction subgraph $G_{i,j}$, and then use a rule-based edge weight function $f$ to assign weights to each edge of the interaction subgraph using multimodal features $H_U$ and $E_V$, obtaining a weighted adjacency matrix $S_{G_{i, j}}$. Then, we use Dijkstra algorithm to obtain top-k shortest paths $P_{i, j}$ as retrieval paths. The formal definition of the process of collecting retrieval paths is as follows:
\begin{equation}
\begin{aligned}
S_{G_{i,j}} = f(G_{i,j},H_U,E_V) \quad P_{i,j} = Dijkstra_{top-k}(u_i,v_j|S_{G_{i,j}})
\end{aligned}
\end{equation}

\textbf{Multimodal features:} Considering the lack of multimodal information on the user side, in this work, we model multimodal representation learning as a behavior sequential recommendation task based on a sequence encoder, and optimize feature representation by training the sequential recommendation task. Assuming that the multimodal features of each item $i_j$ are composed of $E_{v_j}=(E_{v_j}^t,E_{v_j}^v)$, where the former represents textual features and the latter represents visual features. The behavior sequence of user $u_i$ is $S_{u_i}=\{v_1^{(i)},...,v_{n_i}^{(i)}\}$. The sequential recommendation task is to use the multimodal features of each item in the sequence to generate user features $H_{u_i}$ through the sequence encoder $E_{seq}$, and then retrieve the most relevant items for recommendation prediction through vector similarity. The formal definition of the process for retrieving target item in sequential recommendation tasks is as follows:
\begin{equation}
\begin{aligned}
H_{u_i} = E_{seq}(S_{u_i}|E_V) \quad k= argmax_{v_j \in V}H_{u_i}^T \cdot E_{v_j}
\end{aligned}
\end{equation}

\subsection{Multimodal Representation Learning}
In recommendation datasets, there is often rich multimodal information on the item side. However, due to the problem of information imbalance caused by Modality Gaps, the importance of different modality features is not the same. Directly concatenating various modality features will introduce noise, which will affect representation learning.To alleviate the problem of modality imbalance, we propose a sequential modeling scheme based on joint residual encoding for multimodal representation learning.

\subsubsection{Joint Residual Encoding} 
In order to achieve better modality alignment, we use RQ-VAE technology (see Appendix A.1 for technical details) to unify the semantic mapping of information from different modalities through joint residual encoding, and obtain corresponding modality semantic indexes for constructing ID representations. Specifically, we first use the powerful LLM2CLIP as a multimodal encoder to initialize the text features $E_{V}^t \in R^{|V| \cdot D_t}$ and visual features $E_{V}^v \in R^{|V| \cdot D_v}$ of the item. Then, we construct a Residual Quantized Variational AutoEncoder (RQ-VAE) for each modality feature. In each modality $m \in \{t, v\}$, the original embedding $e_j^m$ is encoded as a semantic embedding $z_j^m$, and the codebook index of each layer is obtained as the semantic ID $\{SID_m^1, SID_m^2,...,SID_m^n\}$ through L-layer codebook level residual quantization, where $n$ is the size of each layer's codebook (the number of cluster centers). Represent the quantized embedding of each modality as $\hat{z}_j^m$, The modality representation reconstructed by the decoder is $\hat{e}_j^m$. In terms of training methods, we take into account the correlation between multimodal embeddings within and between modalities. We use the reconstruction loss $\mathcal{L}_{recon}$ and commitment loss $\mathcal{L}_{comit}$ to maintain the uniqueness within each modality; Use InfoNCE loss $\mathcal{L}_{align}$ to align text and visual features between modalities. The loss function of the joint residual encoding module is as follows:
\begin{equation}
\begin{aligned}
&\mathcal{L}_{mutli} = \mathcal{L}_{align} + \sum_{m \in \{t,v\}}(\mathcal{L}_{commit} +\mathcal{L}_{recon}) \\
&\mathcal{L}_{align} = -\frac{1}{|V|}\sum_{j=1}^{|V|}log\frac{exp(sim(\hat{z}_j^t, \hat{z}_j^v)/\tau)}{exp(sim(\hat{z}_j^t, \hat{z}_j^v)/\tau) + \sum_{j'\neq j}exp(sim(\hat{z}_{j'}^t, \hat{z}_{j}^v)/\tau)}\\
&\mathcal{L}_{commit} = \sum_{i=1}^L||sg[r_i^m]-c_i^m||_2^2 + \beta||r_i^m-sg[c_i^m]||_2^2\\
&\mathcal{L}_{recon} = ||\hat{e}_j^m-e_j^m||_2^2
\end{aligned}
\end{equation}

Where $c_i^m$ represents the cluster center indexed in the i-th layer codebook of $RQVAE_{m}$, and $r_i^m$ is the residual vector of that layer. $sg[]$ represents the stop gradient operator.

\subsubsection{User Behavior Sequential Modeling}
Although the joint residual encoding module can balance the gaps between different modalities to obtain multimodal quantized representations, this representation ignores the collaborative information in recommendation behavior and therefore cannot fully reflect user preferences. To this end, we model the user behavior sequence based on the joint residual encoding module, and learn the multimodal features of the user side by training the sequence encoder.

Regarding the initialization of each item feature in the sequence, in order to prevent catastrophic forgetting of the quantization model, We unified the quantized embedding $\hat{z}_v^m$ and the original feature $e_v^m$ mapping to the same dimension and concatenate them together for feature initialization. To enhance the generalization of representation learning, we set the codebook vector $\{c_i^m\}_{i=1}^L$($m \in \{t,v\}$) used for constructing quantized embeddings as an updating parameter that is continuously updated as the model is trained. The formal definition of initializing the item feature $e_v$ is as follow:
\begin{equation}
\begin{aligned}
e_v = concat[f_{mlp}(e_v^m), \hat{z}_v^m]_{m \in \{t,v\}} \quad \hat{z}_v^m = \sum_{i=1}^L c_i^m
\end{aligned}
\label{semantic}
\end{equation}

Model training: For user $u_i$, assuming their interaction sequence is $S_ {u_i}=[v_1, v_2,..., v_k]$, after processed by the sequence encoder, we obtained the user's multimodal representation $h_i$. Then we calculate the recall probability of positive and negative samples through vector dot product and use InfoNCE loss to train the sequence encoder module:
\begin{equation}
\begin{aligned}
\mathcal{L}_{seq}=-log\frac{exp(h_i\cdot e^+)}{exp(h_i\cdot e^+)+\sum_{v\in e^-}exp(h_i\cdot e_v)}
\end{aligned}
\label{semantic}
\end{equation}

\subsection{Retrieval Path Collection}
Considering the poor explainability and unsuitability of graph explanation schemes based on graph neural networks(GNNs) for recommendation tasks, we propose a heuristic search method based on multimodal features for retrieval path collection. The search process is divided into three steps: first, pruning the original graph to obtain a interaction subgraph, and then assigning weights to each edge based on multimodal feature utilization rules on the interaction subgraph; Afterwards, the shortest path algorithm is used to search on the weighted subgraph to obtain the retrieval path. Now, we will take the recommendation behavior for $(u_i, v_j)$ as an example to illustrate the process of obtaining relevant retrieval paths.

\subsubsection{Graph Pruning}
Due to the large amount of redundant information and noise contained in the original interaction graph, we first extract the L-hop neighbors of $u_i$ and $v_j$ to construct a subgraph $G_{(u_i, v_j)}$. In order to obtain a more concise and accurate explanation, we hope to extract the shortest possible path. Nodes with lower popularity in the interaction graph carry less collaborative information and should be discarded. For this purpose, we perform \textbf{k-core pruning} on subgraph $G_{(u_i, v_j)}$ to obtain k-core subgraph $G_{(u_i,v_j)}^k$. The definition of a k-core subgraph is the unique largest subgraph with the minimum node degree k. K-core pruning is a recursive iterative algorithm that ends when there are no states with an in degree less than k in the graph. The pseudocode of the algorithm can be found in Appendix A.2. It is worth noting that since we are looking for the shortest path between $u_i$ and $v_j$, these two points should be preserved when performing k-core pruning.

\subsubsection{Graph Weighting}
For the interaction subgraph $G_{(u_i,v_j)}^k$, we set $G_{(u_i,v_j)}^k$ as a weighted directed graph and use the edge weight function $f_ {edge}$ to score the explainability of each edge. In order to facilitate the application of the shortest path algorithm, we set the lower the score, the stronger the explainability of the edges. Next, we will take the edge pointing from user $u_m$ to item $v_n$ as an example to explain the edge weight calculation rules. Due to our prior knowledge of the multimodal features of each node, for directed edge $(u_m, v_n)$, the closer the feature's similarity between endpoint $v_n$ and the target user node $u_i$, the stronger its explainability. In addition, we also balanced the impact of node popularity on personalized interpretation. Since nodes with high degrees are often generic and have less information, the higher the endpoint node's degree, the poorer the explainability of directed edge $(u_m, v_n)$. The formal definition of edge weight function is as follows:
\begin{equation}
\begin{aligned}
f_{edge}^{G_{(u_i,v_j)}^k}(u_m,v_n) =
\begin{cases}
\log[(2 - Sim(v_n, u_i)) \cdot \deg(v_n)], & \text{if } v_n \neq v_j, \\
1, & \text{if } v_n = v_j.
\end{cases}
\end{aligned}
\end{equation}

\subsubsection{Graph Retrieval}
In order to make the application of retrieval paths to LLM inference simpler and more effective, in addition to maximizing the explainability of nodes on the retrieval path, we also hope that the path is as short as possible to avoid introducing noise. Therefore, we use the shortest path algorithm Dijkstra to find the top-k shortest path with the starting node of $u_i$ and the ending node of $v_j$ as the search path for $(u_i, v_j)$. Here is the length definition of the path $p=[u_i, n_1, n_2,..., n_k, v_j]$:
\begin{equation}
\begin{aligned}
len(p) = \sum_{i=1}^k f_{edge}(n_{i-1},n_i) \text{ where }n_0=u_i
\end{aligned}
\end{equation}

The retrieval path obtained from this not only satisfies simplicity, but also ensures that each node is an entity strongly related to the recommended behavior to be explained, thereby enhancing overall explainability.

\subsection{LLM reasoning Based on Retrieval Path}
After obtaining the retrieval paths $P_{(u_i, v_j)}$ for the recommended behavior $(u_i, v_j)$, we formalize these retrieval paths into a textual description as external knowledge to assist the LLM inference. Considering that each retrieval path contains rich collaborative information, we also designed a trainable graph neural network(GNN) and Mixture-of-Experts model to extract collaborative information from different paths and map it to the semantic space of the LLM. Below is an introduction to the various modules used in the generation phase.

\subsubsection{Graph Encoder}
$p=[u_i, n_1,..., n_k, v_j]$ represents a retrieval path about $(u_i, v_j)$. We select the 1-hop neighbor of each node on the path to form the retrieval subgraph $G_p$ according to the origin graph $G$. Then we use a trainable GNN model as collaborative encoder $E_C$ to encode $G_p$. 
\begin{equation}
\begin{aligned}
h_{G_p}= MeanPool\{E_C(h_n)| n \in V_{G_p}\}
\end{aligned}
\end{equation}

Where $V_{G_p}$ represents the node set of $G_p$, and $h_n$ represents the multimodal representation of node $n$. Here, we transform the multimodal features of all nodes in the retrieval subgraph $G_p$ into a collaborative space through graph convolution and perform mean pooling as the collaborative features of the retrieval subgraph $G_p$. Due to the existence of multiple retrieval paths for $(u_i, v_j)$, we repeat the above encoding operation for each retrieval path in $P_{(u_i, v_j)}$ and concatenate the collaborative features of each retrieval subgraph to obtain the collaborative representation $h_{(u_i,v_j)}$. The relevant formal definition is as follows:
\begin{equation}
\begin{aligned}
h_{(u_i,v_j)}= Concat\{h_{G_p}|p \in P_{(u_i,v_j)}\}
\end{aligned}
\end{equation}

\subsubsection{MoE Adapter}
In order to enable LLM to understand the collaborative information obtained from graph embedding, we introduce a Mixture-of-Expert model as an adapter. In the Mixture-of-Expert architecture, each expert network is represented by a linear layer that captures different semantic dimensions, and then these expert networks are integrated using a learnable gated routing mechanism. This allows the model to adaptively combine different semantic representations encoded by various experts, so as to effectively bridge the gap between behavior aware collaborative relationship signals and textual language signals. The formal definition of Mixture-of-Experts is as follows:
\begin{equation}
\begin{aligned}
&y = \sum_{i=1}^nG(x)_iE_i(x) \\
&e_i=W_i(Dropout(x)+b_i) \\
&G_\sigma (x)=softmax(x\cdot W_g)
\end{aligned}
\end{equation}

For the collaborative expression $h_{(u_i,v_j)}$ of the recommended pair $(u_i,v_j)$, after the transformation of the MOE adapter, we can get the adapted embedding $s_{(u_i,v_j)}$ and use it as a soft prompt for recommendation in the later reasoning process.

\subsubsection{LLM Reasoning}
For the retrieval path set $P_{(u_i,v_j)}$ obtained in the retrieval phase, we express each path $p$ in formal language. Specifically, we use the profile information to express each node on the path, and then the path transfer relationship is expressed as "buys" (user->item) and "bought by" (item->user).The retrieval path can be expressed as: "<User profile> -> buys -> <Item profile> -> bought by -> <User profile> -> ...".We combine the retrieval path as external knowledge with profile as hard prompts. The following is the definition of instruction prompt:

\fbox{
\parbox{0.4\textwidth}{
Given the item title, item profile, user profile and some retrieval paths about the user, please explain why the user would enjoy this item. Item title: \textbf{[Item title]}. Item profile: \textbf{[Item summary]}. User Profile: \textbf{[User summary]}.Here are several related paths which may reflect his/her preference. \textbf{[Retrieval paths]}. Explanations:
}
}

After being processed by the positional embedding layer, we obtained the token embedding representation $\epsilon_{(u_i,v_j)}$ of hard prompt, and we combined the collaborative adaptation embedding $s_{(u_i,v_j)}$ with it as the final input of LLM.

\begin{table*}[ht]
\centering
\setlength{\tabcolsep}{3pt}
\renewcommand{\arraystretch}{1.1}
\caption{Explainability and Stability results on Three Datasets. ↑ means higher is better; ↓ means lower is better. Superscripts “P”, “R”, and “F1” denote Precision, Recall, and F1-Score, respectively. The subscript “std” indicates the standard deviation of each metric. Bold indicates the best results, while underlined denotes the second-best.}
\label{table2}
\resizebox{\textwidth}{!}{  
\begin{tabular}{l|l|ccccccc|cccccc}
\toprule
\multirow{2}{*}{Category} & \multirow{2}{*}{Models} & \multicolumn{7}{c|}{\textbf{Explainability} $\uparrow$} & \multicolumn{6}{c}{\textbf{Stability} $\downarrow$} \\
& & GPT$_\text{score}$ & BERT$^P_\text{score}$ & BERT$^R_\text{score}$ & BERT$^{F1}_\text{score}$ & BART$_\text{score}$ & BLEURT & USR & GPT$_\text{std}$ & BERT$^P_\text{std}$ & BERT$^R_\text{std}$ & BERT$^{F1}_\text{std}$ & BART$_\text{std}$ & BLEURT$_\text{std}$ \\
\midrule
\multicolumn{15}{c}{\textbf{Baby}} \\
\midrule
\multirow{6}{*}{Baseline} 
& NRT          & 22.36 & 0.1515 & 0.0998 & 0.1260 & -5.2312 & -1.0946 & 0.0094 & 17.09 & 0.1525 & 0.1506 & 0.1351 & 1.0305 & 0.2325 \\
& Att2Seq      & 26.19 & 0.1356 & 0.0887 & 0.1119 & -5.2568 & -1.0701 & 0.0437 & 21.14 & 0.1827 & 0.1608 & 0.1465 & 1.0432 & 0.2667  \\
& PETER        & 45.11 & 0.3278 & 0.2305 & 0.2784 & -4.7111 & -0.8737 & 0.5338 & 26.25 & 0.2051 & 0.2103 & 0.1889 & 1.0288 & 0.4010 \\
& PEPLER       & 24.95 & 0.0951 & 0.0134 & 0.0530 & -3.9049 & -1.2336 & 0.6782 & 21.49 & 0.1731 & 0.2077 & 0.1543 & 1.0600 & \textbf{0.1476} \\
& CER        & 46.11 & 0.3154 & 0.2292 & 0.2716 & -4.7099 & -0.8669 & 0.6078 & 26.02 & 0.2045 & 0.2076 & 0.1863 & 1.0304 & 0.4408 \\
& G-Refer    & 82.35 & \underline{0.5924} & \underline{0.5779} & 0.5856 & -3.2977 & -0.0598 & \underline{0.9989} & 9.28 & 0.1172 & 0.1160 & 0.1114 & 0.6103 & 0.2180 \\
\midrule
\multirow{2}{*}{Base LLM}
& Llama2-7b-Chat-hf & 87.29 & 0.4302 & 0.4395 & 0.4354 & -3.5383 & 0.0425 & \textbf{1.0000} & 6.78 & \textbf{0.0928} & \underline{0.1028} & \textbf{0.0886} & 0.5820 & 0.2133 \\
& Llama2-13b-Chat-hf & \textbf{88.82} & 0.4808 & 0.5300 & 0.5057 & \underline{-3.1185} & \underline{0.1776} & \textbf{1.0000} & \underline{6.67} & 0.1290 & \textbf{0.0992} & \underline{0.1058} & \textbf{0.5130} & \underline{0.1873} \\
\midrule
\rowcolor{gray!10}
Ours & MMP-Refer  & \underline{88.57} & \textbf{0.6180} & \textbf{0.5992} & \textbf{0.6091} & \textbf{-2.8699} & \textbf{0.1881} & \textbf{1.0000} & \textbf{6.28} & \underline{0.1132} & 0.1151 & 0.1090 & \underline{0.5277} & 0.2196 \\
\midrule
\multicolumn{15}{c}{\textbf{Sports}} \\
\midrule
\multirow{6}{*}{Baseline} 
& NRT          & 26.62 & 0.0319 & 0.0574 & 0.0410 & -5.6958 & -1.1106 & 0.0595 & 21.12 & 0.3362 & 0.1572 & 0.2165 & 1.0707 & 0.2767 \\
& Att2Seq      & 26.18 & -0.0326 & 0.0489 & 0.0033 & -5.7100 & -1.1307 & 0.0383 & 21.58 & 0.3653 & 0.1521 & 0.2295 & 1.0720 & 0.2312  \\
& PETER        & 43.54 & 0.2579 & 0.1778 & 0.2170 & -5.1567 & -0.8970 & 0.4670 & 25.85 & 0.2159 & 0.1948 & 0.1822 & 1.0723 & 0.3844 \\
& PEPLER       & 28.83 & 0.0828 & 0.0542 & 0.0674 & -3.4767 & -1.2145 & 0.7108 & 24.03 & 0.1353 & 0.2129 & 0.1364 & 1.2216 & \textbf{0.1620} \\
& CER        & 46.15 & 0.2709 & 0.1779 & 0.2236 & -5.1481 & -0.8999 & 0.5122 & 25.76 & 0.2037 & 0.1963 & 0.1778 & 1.0716 & 0.3759 \\
& G-Refer    & 81.47 & \underline{0.6025} & \textbf{0.5806} & \underline{0.5920} & -3.4256 & -0.0741 & \textbf{1.0000} & 11.08 & 0.1205 & 0.1230 & 0.1166 & 0.6094 & 0.2078 \\
\midrule
\multirow{2}{*}{Base LLM}
& Llama2-7b-Chat-hf & \underline{87.61} & 0.4496 & 0.4669 & 0.4587 & -3.5526 & 0.1148 & \textbf{1.0000} & \underline{6.05} & \textbf{0.0973} & \textbf{0.1024} & \textbf{0.0883} & 0.5715 & 0.1932 \\
& Llama2-13b-Chat-hf & \textbf{89.04} & 0.4754 & 0.5459 & 0.5107 & \underline{-3.2138} & \textbf{0.2050} & \textbf{1.0000} & \textbf{6.00} & 0.1377 & \textbf{0.1024} & 0.1102 & \textbf{0.5335} & \underline{0.1781} \\
\midrule
\rowcolor{gray!10}
Ours & MMP-Refer  & 87.23 & \textbf{0.6155} & \underline{0.5744} & \textbf{0.5953} & \textbf{-3.0914} & \underline{0.1195} & \underline{0.9996} & 6.67 & \underline{0.1109} & \underline{0.1147} & \underline{0.1065} & \underline{0.5595} & 0.2427 \\
\midrule
\multicolumn{15}{c}{\textbf{Clothing}} \\
\midrule
\multirow{6}{*}{Baseline} 
& NRT          & 31.09 & 0.1748 & 0.1009 & 0.1381 & -5.4719 & -1.0274 & 0.0026 & 23.78 & 0.1302 & 0.1593 & 0.1289 & 1.0462 & 0.3112 \\
& Att2Seq      & 35.10 & 0.2079 & 0.1209 & 0.1643 & -5.3831 & -1.0108 & 0.0303 & 25.76 & 0.1431 & 0.1683 & 0.1379 & 1.0539 & 0.3145  \\
& PETER        & 50.83 & 0.3489 & 0.2586 & 0.3029 & -4.8712 & -0.7521 & 0.2737 & 26.19 & 0.2019 & 0.2267 & 0.1945 & 1.0913 & 0.5012 \\
& PEPLER       & 35.96 & 0.1332 & -0.0029 & 0.0625 & -3.7008 & -1.2115 & 0.4579 & 27.93 & 0.1663 & 0.2425 & 0.1648 & 0.9730 & \textbf{0.1435} \\
& CER        & 52.89 & 0.3687 & 0.2592 & 0.3129 & -4.8688 & -0.7577 & 0.1790 & 27.50 & 0.2103 & 0.2349 & 0.2031 & 1.1084 & 0.5104 \\
& G-Refer    & 82.39 & \underline{0.5818} & \textbf{0.5544} & \underline{0.5685} & -3.3886 & -0.0256 & \underline{0.9961} & 10.31 & 0.1267 & 0.1248 & 0.1199 & 0.5969 & \underline{0.1747} \\
\midrule
\multirow{2}{*}{Base LLM}
& Llama2-7b-Chat-hf & \underline{87.71} & 0.4241 & 0.4472 & 0.4360 & -3.4871 & 0.0747 & \textbf{1.0000} & \underline{7.31} & \textbf{0.1074} & \underline{0.1097} & \textbf{0.0958} & \underline{0.5474} & 0.2223 \\
& Llama2-13b-Chat-hf & \textbf{88.85} & 0.4365 & 0.5169 & 0.4767 & \underline{-3.1846} & \textbf{0.1553} & \textbf{1.0000} & \textbf{7.05} & 0.1443 & \textbf{0.1049} & 0.1133 & \textbf{0.5049} & 0.1987 \\
\midrule
\rowcolor{gray!10}
Ours & MMP-Refer & 86.78 & \textbf{0.6168} & \underline{0.5473} & \textbf{0.5822} & \textbf{-2.9749} & \underline{0.0822} & \textbf{1.0000} & 7.32 & \underline{0.1136} & 0.1194 & \underline{0.1096} & 0.5478 & 0.2679 \\
\bottomrule
\end{tabular}
}
\end{table*}

\subsubsection{Train and Inference}
We aim to minimize the loss between the prediction probability of the next token in the sequence and the actual next token. Therefore, we use negative log-likelihood (NLL) as the training loss, and the loss function is defined as follows:
\begin{equation}
\begin{aligned}
L = -\frac{1}{N}\sum_{i=1}^N\sum_{c=1}^{C_i}y_{ic}\cdot log(\hat{y}_{ic})
\end{aligned}
\end{equation}

Here, N is the number of explanations, $C_i$ is the character count in each explanation, and $y_{ic}$ and $\hat{y}_{ic}$ represent the actual and predicted tokens, respectively.

As for the ground-truth explanation of the recommendation system, we followed G-Refer to use the profile and the review information in the interaction, and take ChatGPT as the benchmark for generation.

\section{Experiment}
We evaluated our model on real-world datasets to assess its performance in about generation explainability. And answered the following question, \textbf{R1:} How effective is our method compared to other baseline models, whether it is better than the state-of-the-art models currently available; \textbf{R2:} The impact of the main modules of our model on overall performance; \textbf{R3:} The impact of the number of retrieval paths on model performance; \textbf{R4:} Efficiency comparison with traditional graph explanation schemes.

\subsection{Experiment Setting}
\subsubsection{Dataset}
We conducted experiments using three publicly available multimodal recommendation datasets, namely Baby, Sports, and Clothing subsets from the Amazon Review Dataset\footnote{https://jmcauley.ucsd.edu/data/amazon/links.html}. More details about the dataset can be found in Appendix B.1.

\subsubsection{Metrics}
In order to evaluate the explanation performance of the model, we followed G-Refer and adopted a set of metrics aimed at capturing the semantic explainability and stability of generated explanations. Traditional metrics based on n-gram syntax, such as BLUE~\cite{papineni2002bleu} and ROUGE~\cite{lin2004rouge}, are insufficient to achieve their goals due to their inability to fully capture semantic meanings. Specifically, we use GPTscore, BERTscore, BARTscore, BLEURT, and USR to measure explainability. Among them, GPTScore~\cite{wang2023chatgpt} maintains consistency with human judgment by comparing the semantic similarity between generated explanations and ground truth explanations; BERTScore~\cite{zhang2019bertscore} utilizes BERT's context embedding to calculate token level cosine similarity; BARTScore~\cite{yuan2021bartscore} utilizes the BART model to conceptualize evaluation as a text generation task, assigning scores based on the probability of regenerating reference text; BLEURT~\cite{sellam2020bleurt} adopts a novel synthetic data pre-training method to enhance generalization; USR~\cite{li2021personalized} measures the uniqueness of generated explanations by calculating the ratio of unique sentences to total sentences. To further evaluate the quality stability, we analyzed the standard deviation of these scores, where lower values indicate more consistent performance. 

\subsubsection{Baselines}
We introduce six state-of-the-art baselines, including NRT, Att2Seq, PETER, PEPLER, CER, and G-Refer. More details of the compared baselines can be found in Appendix B.2.

\subsubsection{Implementation Details}
All experiments were conducted on a 4 NVIDIA RTX-4090 GPUs. Regarding the selection of the model base, we use SASRec as the sequential recommendation model in multimodal representation learning and use the llama2-3b-hf model for explanation generation. Regarding the multimodal representation learning module, the embedding dimension of LLM2CLIP is 1280; The codebook size of RQVAE is 256, the number of codebooks is 4, and the encoder output dimension and codebook tensor dimension are 64; The hidden vector dimension of the sequence encoder is 64. Regarding the retrieval path module, when pruning the entire graph to obtain interactive subgraphs, all datasets default to 3-hop, 2-core pruning. Regarding the LLM inference module, the graph encoder defaults to using the GAT model, and we have also discussed the results of using other models for graph encoding. Set the fine-tuning epoch to 3, the learning rate to 1e-5, and the maximum length to 2048.

\subsection{Model Performance(R1)}
To demonstrate the superiority of our approach in recommendation explainability, we conducted comparative analysis on three datasets and six baseline models, and the results are shown in Table ~\ref{table2}. According to observations, MMP-Refer has shown excellent performance in both explainability and stability. Compared with the baseline model G-Refer, which has the best overall performance, our model has improved the $Bert_{score}^{F1}$ metric by 4.01\%, 0.56\%, and 2.41\% on the three datasets, and by 12.9\%, 9.75\%, and 12.2\% on the $Bart_{score}$ metric, respectively. This indicates that our work, due to the additional introduction of multimodal information, can better utilize item features to mine user preferences and provide personalized explanations. And in terms of stability indicators, our method is also superior to the G-Refer method, which can be attributed to the fact that our retrieval path is based on multimodal feature utilization rule heuristic search, and compared to directly using graph explanation methods, its generalization is stronger. Finally, considering that most baseline models use traditional deep learning methods, in order to more intuitively demonstrate the effectiveness of our approach, we conducted comparative tests with backbone models of different parameter sizes in the same series. The experimental results showed that our method has superior overall explainability compared to the 7B model, but slightly lower stability; Compared with the 13B model, it still demonstrates advantages in explainability and has stronger stability. This fully demonstrates the effectiveness of our framework. In addition, we also employed human evaluation. Please refer to Appendix C for details.

\subsection{Ablation Study(R2)}

\begin{table*}[htbp]
\centering
\caption{Various Ablation Experiment Results on Baby and Sport Dataset.}
\label{table3}
\renewcommand{\arraystretch}{1.2}
\setlength{\tabcolsep}{6pt}
\begin{threeparttable}
\begin{tabular}{c|cccc|cccc}
\toprule
\textbf{Datasets} & \multicolumn{4}{c|}{\textbf{Baby}} & \multicolumn{4}{c}{\textbf{Sports}}  \\
\midrule 
\textbf{Ablation}  & 
\textbf{BERT$^{F1}_{score}\uparrow$} & \textbf{BART$\uparrow$} &
\textbf{BERT$^{F1}_{std}\downarrow$} & \textbf{BART$_{std}\downarrow$} & \textbf{BERT$^{F1}_{score}\uparrow$} & \textbf{BART$\uparrow$} &
\textbf{BERT$^{F1}_{std}\downarrow$} & \textbf{BART$_{std}\downarrow$} \\
\midrule
full training & 0.6091 & -2.8699 & 0.1090 & 0.5277 & 0.5953 & -3.0914 & 0.1065 & 0.5595 \\
\midrule
\multicolumn{7}{l}{\textbf{The Effect of Various GNN Encoder}} \\
\quad w/GCN & 0.6074 & -2.9122 & 0.1098 & 0.5308 & 0.5921 & -3.0873 & 0.1026 & 0.5447 \\
\quad w/GT & 0.6067 & -2.8898 & 0.1105 & 0.5354 & 0.5935 & -3.0836 & 0.1042 & 0.5530 \\
\quad w/o graph-level  & 0.6058 & -2.9356 & 0.1095 & 0.5260 & 0.5905 & -3.0978 & 0.1049 & 0.5550\\
\midrule
\multicolumn{7}{l}{\textbf{The Effect of Retrieval Path}} \\
\quad w/o Lora & 0.4730 & -3.4203 & 0.1264 & 0.5349 & 0.4873 & -3.5330 & 0.1135 & 0.5460 \\
\midrule
\multicolumn{7}{l}{\textbf{The Effect of Joint Residual Encoding}} \\
\quad w/o Multi-SID & 0.6039 & -2.9474 & 0.1085 & 0.5320 & 0.5942 & -3.1114 & 0.1050 & 0.5527 \\
\bottomrule
\end{tabular}
\end{threeparttable}
\end{table*}

\subsubsection{The Effect of Various GNN Encoder}
We investigated how collaborative representations of retrieval paths derived from different GNN encoders affect model performance. As shown in Table ~\ref{table3}, the model achieves competitive explainability across various recommendation scenarios and GNN architectures, demonstrating the robustness of our multimodal representation learning. Mapping retrieval paths into a shared collaborative space enables different graph encoders to extract effective collaborative representations. When the collaborative features are removed and only textual features are used for LLM fine-tuning, explainability declines, suggesting that collaborative subgraph information enhances the model’s understanding of user intent. Nevertheless, the performance gain remains limited, partly due to the reduced attention to soft prompt labels with longer inputs and the difficulty of further improvement given the LLM’s limited parameter capacity, even with rich retrieval path information.

\subsubsection{The Effect of Retrieval Path}
We explored whether there is a significant improvement in the explainability performance of the frozen LLM by relying solely on text and collaborative information from retrieval paths. Combining Table ~\ref{table2} and Table ~\ref{table3}, when we remove the Lora fine-tuning and combine it with the retrieval path, the explainability performance of the 3B model is better than that of the 7B model in the same series and closer to the 13b model. This further demonstrates the effectiveness of our retrieval path collection scheme.

\subsubsection{The Effect of Joint Residual Encoding}
We explored the impact of the joint residual encoding module on multimodal representation learning, as shown in Table ~\ref{table3}. When we remove the joint residual encoding module, the model's explainability will decrease to a certain extent. This is because after removing the quantization embedding, the user sequential modeling will only rely on the original features during feature initialization. Due to the lack of modality alignment, the final multimodal features will be more unevenly distributed, making the representation unable to reflect the user's personalized information well, ultimately affecting the explainability. Detailed arguments can be found in Appendix D.

\subsection{Hyperparameter Study(R3)}

\begin{figure}[t]
    \centering
    \includegraphics[width=1\linewidth]{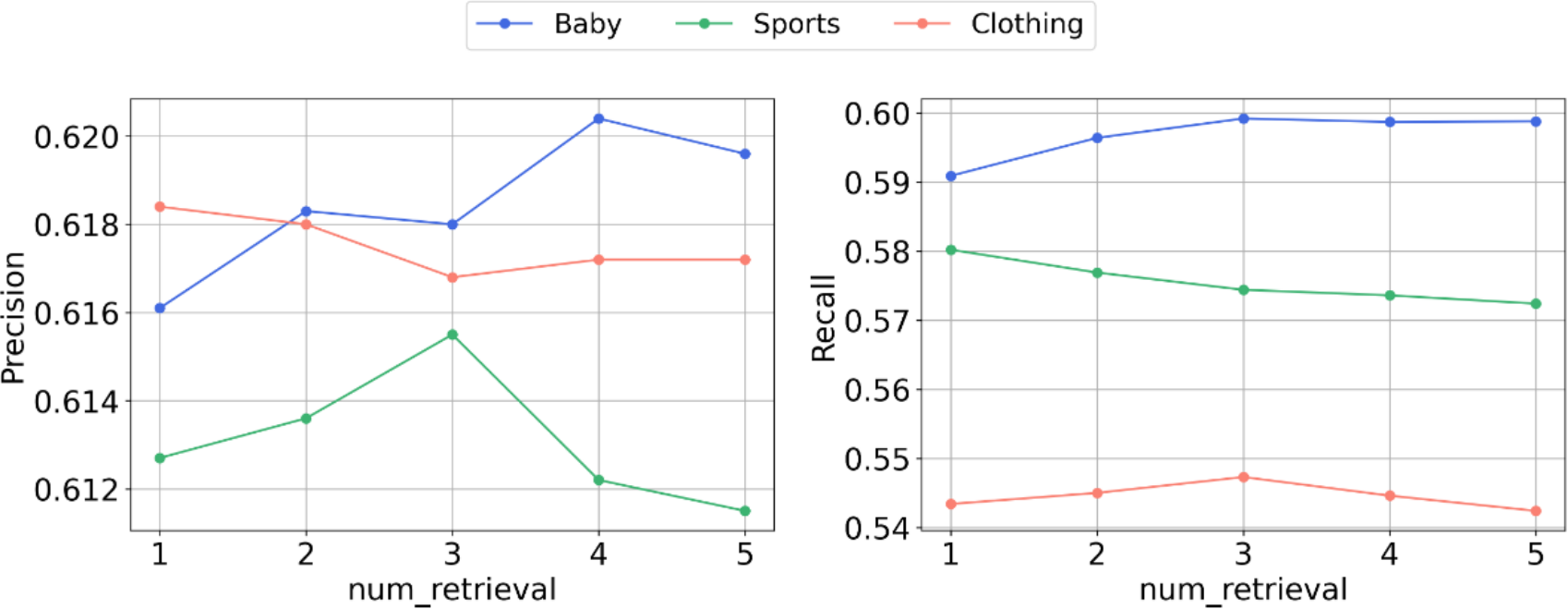}
    \caption{Performance of different retrieved number}
    \label{hyper}
\end{figure}

We discussed the impact of the number of retrieval paths k (from 1 to 5) on the model's explanation performance (precision $Bert_{score}^P$ and recall $Bert_{score}^R$). From the figure ~\ref{hyper}, it can be seen that the model explanation performance reaches its optimal level for all datasets when k=3. This is because when the number of retrieval paths is small, the profile information of the nodes along the retrieval path may not fully explain the user's preference information; When there are too many retrieval paths, noise information will be introduced to the LLM judgment, thereby reducing performance.

\subsection{Efficiency Analysis(R4)}
In Table ~\ref{table4}, we compared the computational efficiency of our work with the PageLink~\cite{zhang2023page} method used by G-Refer in collecting retrieval paths. For the sake of fairness, all retrieval processes are performed on the CPU. It can be seen that our method has increased efficiency by 3 times, 4 times, and 6.06 times on the three datasets, respectively. Moreover, the sparser the recommended dataset, the more significant the efficiency improvement of our method.

\begin{table}[htbp]
  \centering
  \caption{Comparison of Retrieval Time Efficiency.}
  \label{table4}
  \begin{tabular}{ccccc}
    \toprule
    \textbf{Datasets} & \textbf{Graph Num} & \textbf{Scheme} & \textbf{Time} & \textbf{Speed-up Ratio} \\
    \midrule
    \multirow{2}{*}{Baby} & \multirow{2}{*}{14833} & Ours & 1.5h & \multirow{2}{*}{$\mathbf{3.0\times}$} \\
    & & Page-Link & 4.5h & \\
    \cline{1-5} 
    \multirow{2}{*}{Sports} & \multirow{2}{*}{13700} & Ours & 1.0h & \multirow{2}{*}{$\mathbf{4.0\times}$} \\
    & & Page-Link & 4.0h & \\
    \cline{1-5}
    \multirow{2}{*}{Clothing} & \multirow{2}{*}{13166} & Ours & 0.33h & \multirow{2}{*}{$\mathbf{\approx 6.06\times}$} \\
    & & Page-Link & 2.0h & \\
    \bottomrule
  \end{tabular}
\end{table}
\section{Conclusion}
We propose a LLM-based explainable recommendation framework MMP-Refer based on multimodal retrieval paths. The innovation of our work mainly lies in two aspects: firstly, we propose a user sequential modeling method based on joint residual encoding to introduce multimodal features while balancing modality gaps; One is to design a simple and efficient heuristic search method to collect retrieval paths for explainable recommendations. Comprehensive experiments have verified the effectiveness of our model.
\bibliographystyle{ACM-Reference-Format}
\bibliography{reference}

\appendix

\section{Technique Details}
\subsection{RQ-VAE}
Residual Quantized Variational Autoencoder (RQ-VAE) aims to tokenize and generate the semantic IDs of the original embedding in a hierarchical manner. Specifically, For the original feature $e$, RQ-VAE first encodes it as a hidden vector $z$ by encoder, and at each layer $h$, we have a codebook $C^h = \{v_k^h\}_{k=1}^K$, and each codebook tensor $\{v_k^h\}$ represents a learnable clustering center. The process of residual quantization is shown in Equ. ~\ref{rqvae}, which first calculates the distance between the residuals and each clustering center in the codebook to find the nearest clustering center, and the distance between the two is the next level of residuals.
\begin{equation}
\label{rqvae}
\begin{aligned}
c_i = arg&min_k||r_i-v_k^i||_2^2 \\
r_{i+1} &= r_i - v_{c_i}^i
\end{aligned}
\end{equation}

After obtaining the H-layer codebook and the corresponding residuals, sum the residuals of each layer to get the quantization result $\hat{z}$, i.e., $z=\sum_{i=1}^H r_i$, then use $\hat{z}$ as the decoder input to reconstruct the item embedding $\hat{e}$, and the loss of the whole process is as follow:
\begin{equation}
\begin{aligned}
\mathcal{L}_{RECON}&=||e-\hat{e}||_2^2 \\
\mathcal{L}_{RQ}=\sum_{i=1}^H||sg[r_i]&-v_{c_i}^i||_2^2+\beta||r_i-sg[v_{c_i}^i]||_2^2 \\
\mathcal{L}_{RQ-VAE}&=\mathcal{L}_{RECON}+\mathcal{L}_{RQ}
\end{aligned}
\end{equation}

where $\hat{e}$ is the output of the decoder, $sg[]$ represents the stop-gradient operator, and $\beta$ is a loss coefficient, usually set to 0.25. The overall loss is divided into two parts, $\mathcal{L}_{RECON}$ is the reconstruction loss, and $\mathcal{L}_{RQ}$ is the RQ loss used to minimize the distance between cluster center vectors and residual vectors.

\subsection{k-core Pruning}
K-core pruning is a graph reduction method that iteratively removes nodes with degrees lower than a given threshold $k$, retaining only the subgraph where all nodes satisfy $deg(v)\geq k$. In Alg. ~\ref{k-core}, node degrees are initialized, and nodes with insufficient degrees are enqueued and removed along with their edges. Neighbor degrees are updated accordingly, and the process continues until no nodes violate the degree constraint, yielding the final k-core subgraph.

\begin{algorithm}
\caption{K-Core Pruning}
\label{k-core}
\begin{algorithmic}[1]
\Require Graph $G=(V,E)$, integer $k$
\Ensure Subgraph $G'=(V',E')$ where $\forall v\in V': \deg_{G'}(v)\ge k$
\State $deg[v] \gets$ degree of node $v$ in $G$ for all $v\in V$
\State Initialize empty queue $Q$
\For{each $v\in V$}
    \If{$deg[v] < k$}
        \State enqueue $v$ into $Q$
    \EndIf
\EndFor

\State $V' \gets V$, $E' \gets E$
\While{$Q$ is not empty}
    \State $u \gets$ dequeue($Q$)
    \If{$u \notin V'$} 
        \State \textbf{continue}
    \EndIf
    \State Remove $u$ from $V'$
    \For{each neighbor $v$ of $u$ in current $G'$}
        \State Remove edge $(u,v)$ from $E'$
        \State $deg[v] \gets deg[v] - 1$
        \If{$deg[v] < k$}
            \State enqueue $v$ into $Q$
        \EndIf
    \EndFor
\EndWhile

\State \Return $G'=(V',E')$
\end{algorithmic}
\end{algorithm}

\section{More Experiment Details}
\subsection{Details of Datasets}
Table ~\ref{table1} displays the statistical data for each dataset, where 'Train ($u$-$i$)' and 'Test ($u$-$i$)' represent the number of explanation behavior pairs in the training and testing sets. We use GPT-3.5-Turbo to generate configuration information and standard explanations for users and items based on metadata and review information.

\begin{table}[ht]
\centering
\caption{Dataset Statistics}
\label{table1}
\resizebox{0.5\textwidth}{!}{  
\begin{tabular}{lrrrrr}
\toprule
\textbf{Dataset} & \textbf{\# Users} & \textbf{\# Items} & \textbf{\# Interactions} & \textbf{\# Train ($u$-$i$)} & \textbf{\# Test ($u$-$i$)} \\
\midrule
Baby & 19445 & 6956 & 159624 & 11870 & 2963 \\
Sports & 35598 & 18267 & 295366 & 10958 & 2742 \\
Clothing & 39387 & 22478 & 270806 & 10529 & 2637 \\
\bottomrule
\end{tabular}
}
\end{table}

\subsection{Details of Baselines}
We compare our model’s performance against the following baselines:
\begin{itemize}[leftmargin=*, noitemsep] 
\item \textbf{Att2Seq~\cite{dong2017learning}} : Utilizes an attention-based attribute-to-sequence model to generate reviews based on attribute information.
\item \textbf{NRT~\cite{li2017neural}} : Predicts ratings and generates abstractive tips for recommendations using multi-task learning to optimize parameters.
\item \textbf{PETER~\cite{li2021personalized}} : PETER is a personalized transformer model for explainable recommendation. It maps user and item IDs to the generated explanation text, connecting the IDs and words.
\item \textbf{PEPLER~\cite{li2023personalized}} : PEPLER leverages pretrained transformer to generate explainable recommendations based on prompts that incorporate user and item ID vectors. There are several variants of PEPLER, in our experiment, we chose to use the continuous prompt learning version.
\item \textbf{CER~\cite{raczynski2023problem}} : Estimates the discrepancy between predicted ratings and explanation-based ratings to enhance rating-explanation coherency.
\item \textbf{G-Refer~\cite{li2025g}}: Utilize Graph Retrieval-augmented large language models for explainable recommendation.
\end{itemize}

\textbf{Replication Statement:}
Here is an additional explanation of the replication method for each baseline model. For the Att2Seq, NRT, PETER, PERLER, and CER models, we used the Sentires Guide tool to extract sentiment quadruples from user review information in the original dataset, and trained and tested them using the same dataset segmentation method as the original paper. For the G-Refer model, we used the same training and testing samples as our own model, constructed an interaction graph based on the interaction information of the original dataset, and performed link prediction and retrieval path search generation on the edges corresponding to each sample in the interaction graph, rigorously reproducing the paper.

\section{Human Evaluation}

\begin{figure}[t]
    \centering
    \includegraphics[width=\linewidth]{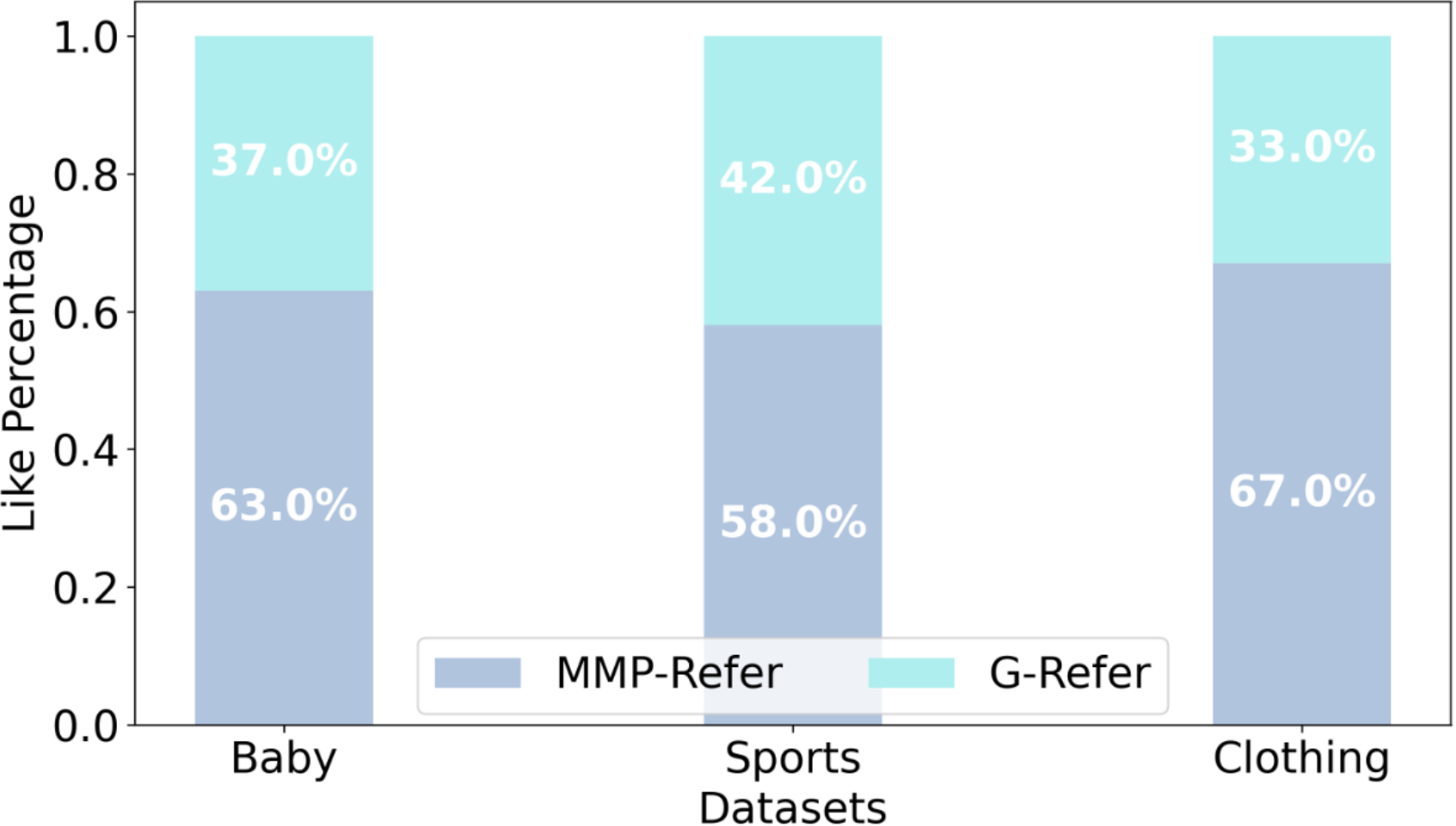}
    \caption{Human evaluation comparing MMP-Refer with G-Refer}   
    \label{fig2}
\end{figure}

In addition to using tools such as BERT and BLEURT, we also employed manual evaluation to assess the explainability of the model. We randomly selected 200 samples from the test set section of each dataset, and then used our own model and G-Refer model with the best text generation ability in the baseline to generate explanations for each sample. Then we designed a Single choice question survey and distributed it to five researchers, asking them to make the best choice. In the end, we synthesized the choices of each researcher and presented the results of manual evaluation in Fig \ref{fig2}. It can be seen from this that the explanations generated by the MMP-Refer model are more likely to be favored. We believe this is due to the introduction of multimodal knowledge through retrieval paths during the fine-tuning process of the model, resulting in more accurate and personalized insights.

\begin{table*}[ht]
    \centering
    \caption{A case from Baby Dataset, which proved the effectiveness of the retrieval path and powerful explanation performance of the MMP-Refer.}
    \label{case}
    \begin{tabularx}{\textwidth}{
        >{\centering\arraybackslash}m{2.5cm} 
        |                   
        >{\raggedright\arraybackslash}X      
    }
        \toprule 
        \multicolumn{2}{c}{\textbf{Inputs}} \\ 
        \midrule 

        \textbf{User Profile} & 
        This user is likely to enjoy \textcolor{red}{practical} and \textcolor{red}{colorful} toddler feeding accessories that enhance mealtime experiences, as well as items that are designed for \textcolor{red}{convenience} and \textcolor{red}{safety} for young children. \\
        \cmidrule{1-2} 

        \textbf{Item Profile} & 
        Parents looking for \textcolor{blue}{colorful}, leak-proof baby bottles that are \textcolor{blue}{easy} for babies to grip and hold, with quality nipples and rings, would enjoy the Nuby 3 Pack Tinted Bottle. \\
        \cmidrule{1-2}

        \textbf{Shortest Retrieval Path} &
        \makecell[l]{
            \textbf{Node 1(item):} Parents looking for a \textcolor{blue}{convenient} and portable utensil set for their babies or toddlers, with \textcolor{blue}{a variety} \\ \textcolor{blue}{of colors} and utensils, would enjoy this Nuby 9 Piece Fork and Spoon Travel Set. \\
            \textbf{Node 2(user):} This user is likely to enjoy items that are \textcolor{red}{practical} for parents with young children, focusing on \\ \textcolor{red}{convenience}, \textcolor{red}{safety}, and comfort, such as feeding tools, storage solutions, and play mats. \\
            \textbf{Node 3(item):} Parents with babies looking for a \textcolor{blue}{safe}, portable crib for travel would enjoy this item, particularly \\ those concerned about non-toxic materials and \textcolor{blue}{ease} of setup. \\
            \textbf{Node 4(user):} The user is likely to enjoy baby and infant care items such as bottles, crib sheets, swaddles, and \textcolor{red}{safety} \\ products. They prefer \textcolor{red}{practical} items that are easy to use, \textcolor{red}{colorful}, and affordable, but might have preferences \\ for less girly colors for certain products. \\
        } \\
        
        \midrule 
        \multicolumn{2}{c}{\textbf{Explanations}} \\ 
        \midrule 
        
        \textbf{Ground Truth} &
        The user would enjoy the Nuby 3 Pack Tinted Bottle for their toddler because it is colorful, leak-proof, easy for babies to grip, and has quality nipples, making mealtime fun, safe, and convenient. \\
        \cmidrule{1-2}
        
        \textbf{MMP-Refer} &
        The user would enjoy the Nuby 3 Pack Tined Bottle for its colorful design, leak-proof quality, and convenience, aligning perfectly with their preference for practical and colorful toddler feeding accessories. \\

        \bottomrule 
    \end{tabularx}
\end{table*}

\section{More Analysis about Joint Residual Encoding}
\begin{figure}[t]
    \centering
    \includegraphics[width=\linewidth]{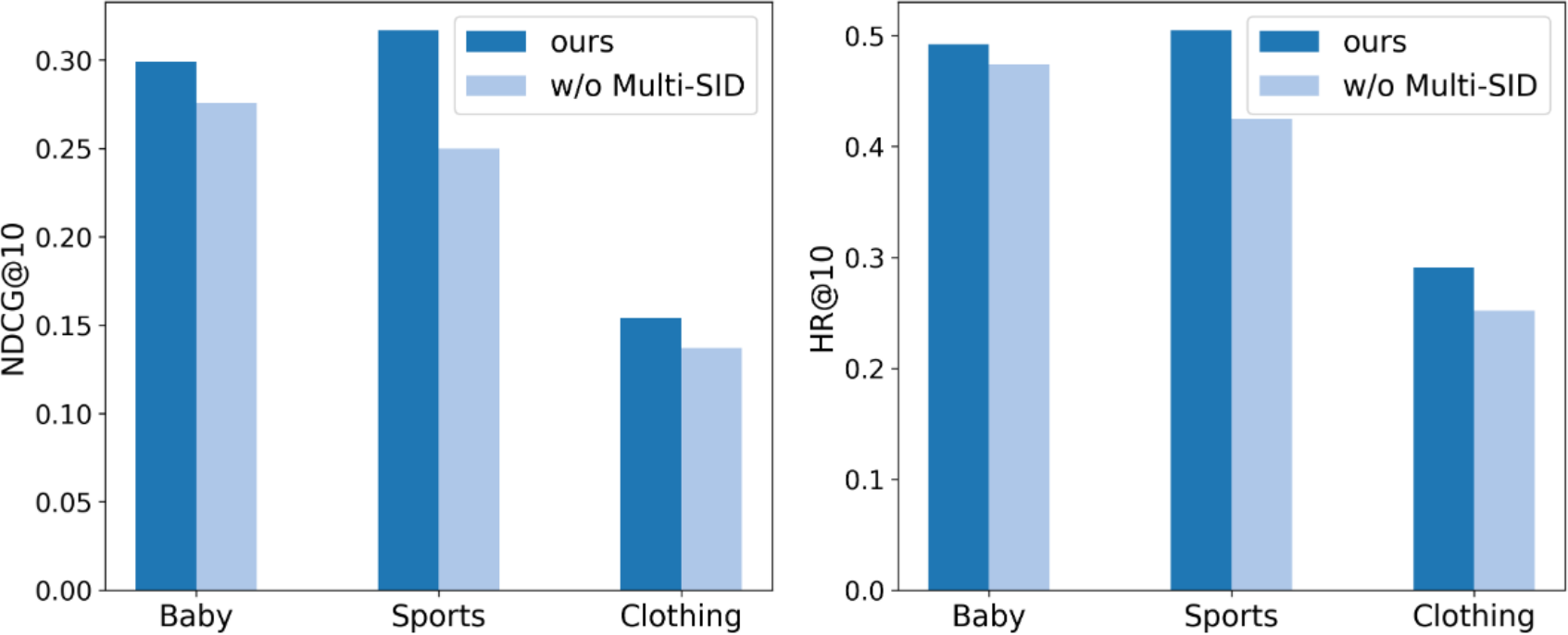}
    \caption{Ablation study of Joint Residual Encoding on the sequential recommendation performance}   
    \label{fig5}
\end{figure}

We further explored the impact of joint residual encoding module on multimodal representation learning from two aspects: sequential recommendation performance and feature distribution. As shown in Figure ~\ref{fig5}, after removing the quantized embedding, the recommendation performance of the sequential recommendation model decreased on all three datasets, which further indicates that multimodal representation cannot reflect user recommendation preferences well, thereby affecting the overall explanation performance.

\begin{figure}[t]
    \centering
    \includegraphics[width=\linewidth]{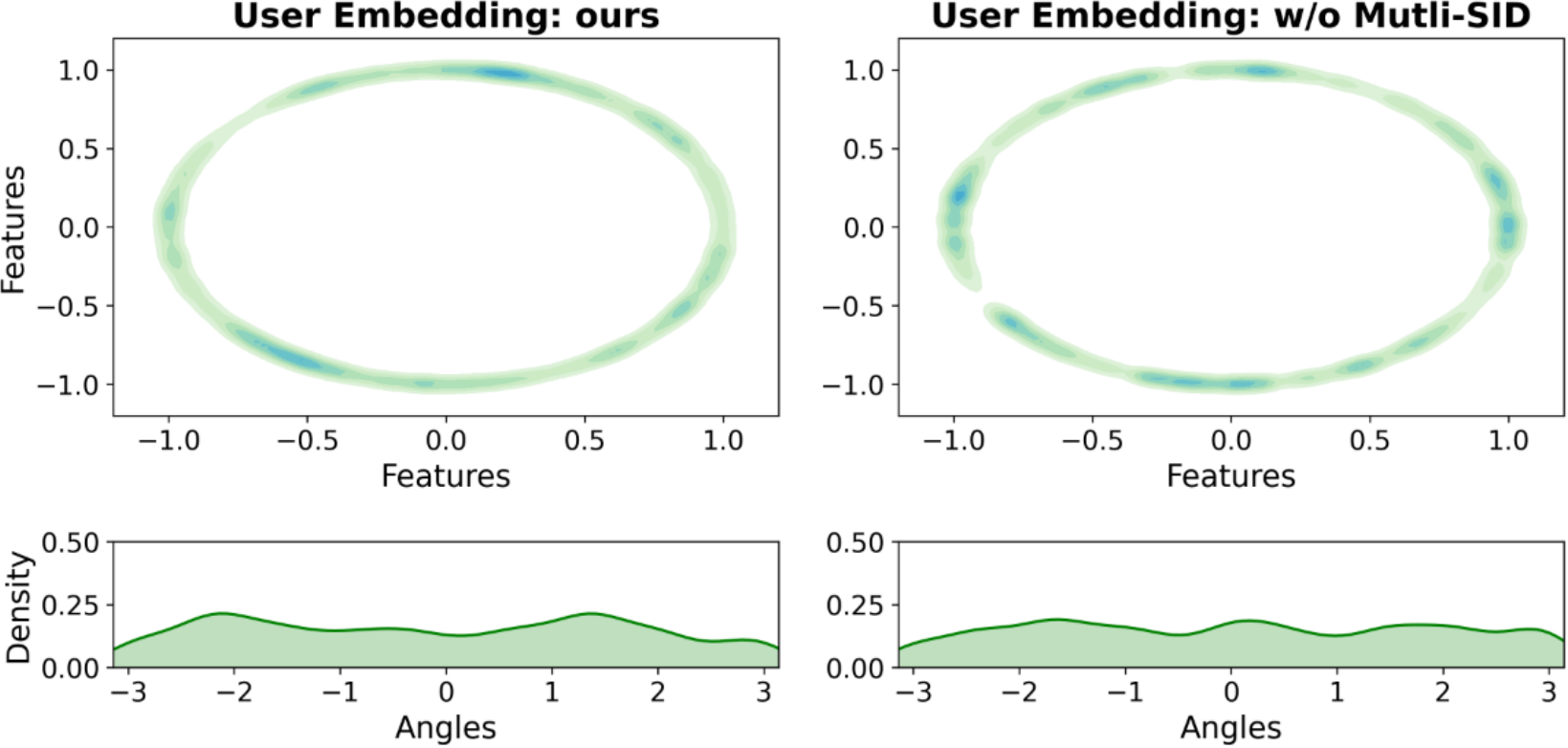}
    \caption{The distribution of representations on the user side.The left of the figure shows the distribution of full training features, while the right displays the distribution without joint residual encoding.}   
    \label{fig6}
\end{figure}

\begin{figure}[t]
    \centering
    \includegraphics[width=\linewidth]{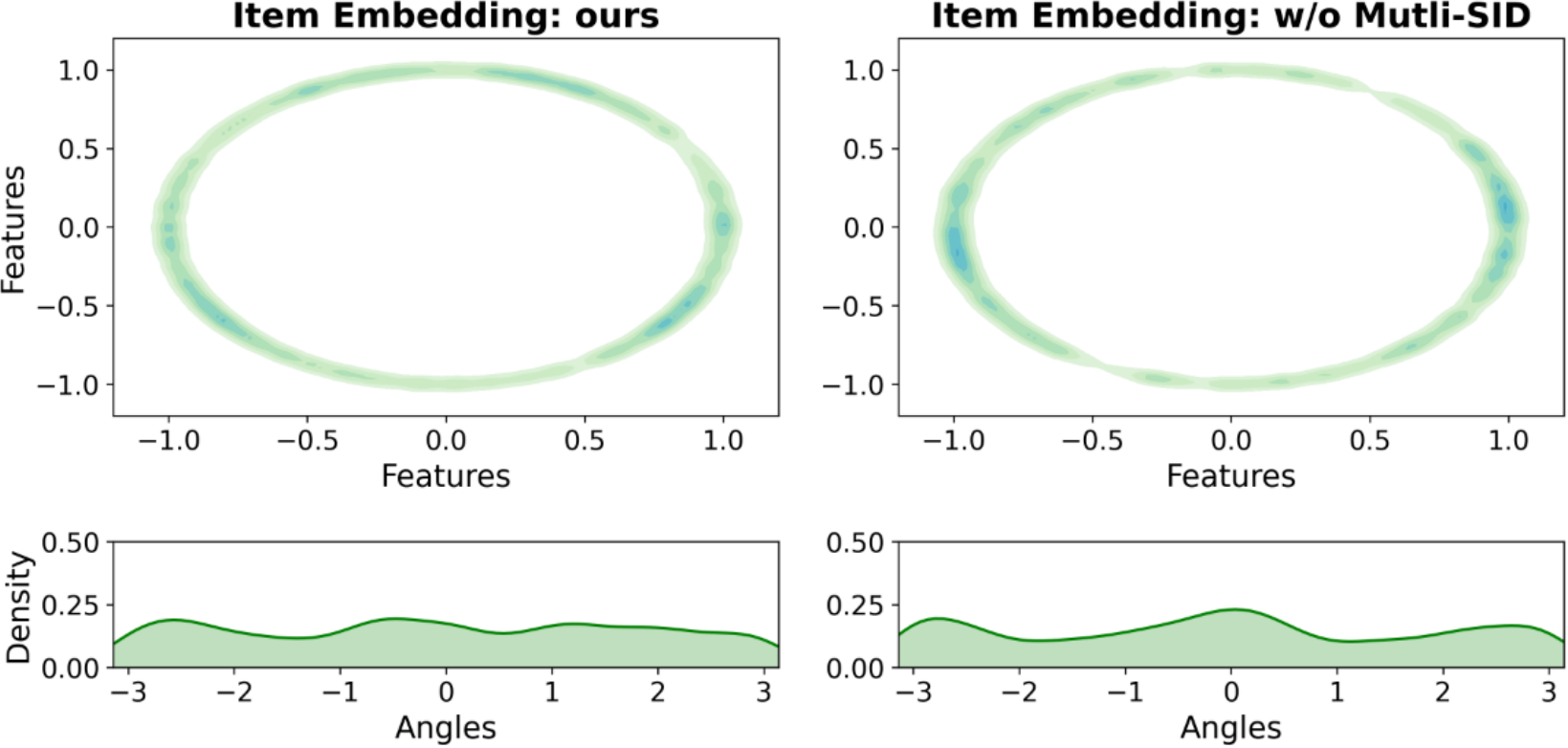}
    \caption{The distribution of representations on the item side.The left of the figure shows the distribution of full training features, while the right displays the distribution without joint residual encoding.}   
    \label{fig7}
\end{figure}

In order to better understand the contribution of joint residual encoding to balancing modality gaps, we visualized the feature distributions of users and items on the Sports dataset after sequential modeling training. Specifically, we randomly selected 5000 users and 3000 items from the Sports dataset and mapped their representations to a two-dimensional space using T-SNE. Next, we use Gaussian kernel density estimation to plot the two-dimensional feature distribution. In short, the embedding representation of the mapping indicates that the closer the hypersphere is to a circular ring or the smoother the feature distribution, the closer the distribution is to a uniform distribution. From Figures ~\ref{fig6} and ~\ref{fig7}, it can be seen that after removing the joint residual encoding module, both user features and item features show obvious community clustering and uneven unipeaks in kernel density estimation, indicating poor distinguishability of features and inability to avoid noise caused by modality gaps, thereby reducing recommendation performance.

\section{Case Study}
We presented an example to further illustrate the effectiveness of the retrieval path and demonstrate the powerful explanatory power of the MMP-Refer model. From Table ~\ref{case}, it can be seen that the configuration information of each node in the retrieval path is highly correlated with the target user and item. The retrieval path model further enhances the perception of user preferences and item attributes, and the final explanation provided is highly similar to the standard answer, which proves the effectiveness of our solution.









\end{document}